\documentclass[12pt,cmcyralt,epsfig]{article}

\usepackage{epsfig}

\newcommand{\nn}{\nonumber}
\newcommand{\beq}{\begin{equation}}
\newcommand{\eeq}{\end{equation}}
\newcommand{\ba}{\begin{eqnarray}}
\newcommand{\ea}{\nonumber \end{eqnarray}}
\newcommand{\be}{\begin{eqnarray}}
\newcommand{\ee}{\nonumber \end{eqnarray}}

\newcommand{\bs}{\begin{slide}}
\newcommand{\es}{\end{slide}}
\newcommand{\bc}{\begin{center}}
\newcommand{\ec}{\end{center}}
\newcommand{\bi}{\begin{enumerate}}
\newcommand{\ei}{\end{enumerate}}

\def\fun#1#2{\lower3.6pt\vbox{\baselineskip0pt\lineskip.9pt
\ialign{$\mathsurround=0pt#1\hfil##\hfil$\crcr#2\crcr\sim\crcr}}}

\begin{document}

\title{Observation of
 tensor glueball in the reactions $p\bar
p\to\pi\pi,\eta\eta,\eta\eta'$ }

\author{ V.V. Anisovich and
A.V. Sarantsev\\  Petersburg Nuclear Physics Institute, Gatchina
188300, Russia}

\date{19.03.05}

\maketitle

\begin{abstract}
Partial wave analysis of the reactions $p\bar
p\to\pi\pi,\eta\eta,\eta\eta'$ in the region of invariant masses
1900--2400 MeV indicates to the existence of
four relatively narrow
tensor-isoscalar resonances
$f_2(1920)$, $f_2(2020)$, $f_2(2240)$, $f_2(2300)$ and the
broad state $f_2(2000)$. The determined decay couplings
of the broad resonance
$f_2(2000)\to\pi^0\pi^0,\eta\eta, \eta\eta'$ satisfy the relations
appropriate to those of tensor glueball, while the couplings of other
tensor states do not, thus verifying the glueball nature of
$f_2(2000)$.

\end{abstract}

PACS numbers: 14.40-n, 12.38-t, 12.39-MK

In \cite{Ani}, the combined partial wave analysis was performed for
the high statistics data on the reactions
 $p\bar p\to  \pi^0\pi^0,\eta\eta,\eta\eta'$
taken  at antiproton momenta 600, 900, 1150, 1200, 1350, 1525, 1640,
1800 and 1940 MeV/c
together with  data obtained for polarised target in the reaction
$\bar p p\to \pi^+\pi^-$ \cite{Eisen}
that resulted in the determination of  a number of  isoscalar
resonances $f_J$ with $J=0,2,4$ (for the review see
\cite{PNPI,ufn04,Bugg}).  In the $02^{++}$-sector,
 five states are required to describe
the data \cite{Ani,PNPI}:
\be
\label{1}
{\rm Resonance}  &  {\rm Mass (MeV)} & {\rm  Width (MeV)}  \\
 f_2(1920) & 1920\pm 30 & 230\pm 40\nn \\
 f_2(2000) &  2010\pm 30 &
 495\pm 35 \nn\\
f_2(2020) & 2020\pm 30 & 275\pm 35\nn \\
f_2(2240) & 2240\pm 40 & 245\pm 45\nn \\
f_2(2300) & 2300\pm 35 & 290\pm 50\, .
\ee
The resonance $f_2(1920)$ was observed earlier in spectra
$\omega\omega$ \cite{VES-f2,GAMS-f2,WA} and $\eta\eta'$
\cite{GAMS-eta,WA-eta}, see also compilation \cite{PDG}.
For the broad tensor-isoscalar
resonance in the region around 2000 MeV the recent analyses give:
$M=1980\pm20\,$MeV, $\Gamma=520\pm50\,$MeV in $pp\to pp\pi\pi\pi\pi$
\cite{Bar} and
$M=2050\pm30\,$MeV, $\Gamma=570\pm70\,$MeV in $\pi^-p\to\phi\phi n$
\cite{LL}.
Following  \cite{Ani,Bar,LL} , we denote the broad resonance as
$f_2(2000)$.

The description of data in the reactions
 $p\bar p\to  \pi^0\pi^0,\eta\eta,\eta\eta'$
 is illustrated by Fig. 1. In Fig. 2,3, one can see differential cross
sections
 $p\bar p\to  \pi^+\pi^-$, while Fig. 4 presents the polarisation data.
 In Fig. 5, we show cross sections for $p\bar p\to
\pi^0\pi^0,\eta\eta,\eta\eta'$ in the $^3P_2\bar p p$ and $^3F_2\bar p
p$ waves (dashed and dotted curves) and total $(J=2)$ cross section
(solid curve) as well as the Argand-plots for the $^3P_2$ and $^3F_2$
wave amplitudes at invariant masses $M=1.962$, $2.050$, $2.100$,
$2.150$, $2.200$, $2.260$, $2.304$, $2.360$, $2.410$ GeV.

Partial wave analysis \cite{Ani,PNPI} together with recent
data for
$\gamma\gamma\to K_SK_S$ \cite{L3} and re-analysis
of $\phi\phi$-spectra \cite{LL} have clarified the
situation with $f_2$-mesons in the mass region $1700-2400\,$MeV.
Based on these data, there was performed in \cite{glue-2}  a
systematisation of the  non-exotic $f_2$-mesons  on the
$(n,M^2)$-trajectories, where $n$ is the radial quantum number of the
$q\bar q$-state. The systematisation \cite{glue-2} shows us that
 the broad resonance
$f_2(2000\pm30)$ is an extra state for the
$(n,M^2)$-trajectories being apparently the lowest tensor glueball.
However, the
statement about glueball nature of $f_2(2000)$ was based on indirect
arguments:\\
(i) The leading Pomeron trajectory
 $\alpha_P(M^2)=\alpha_P(0)+\alpha'_P(0)M^2$
has the following values for the intercept  and slope : $\alpha
(0)\simeq 1.10 - 1.30$ and $\alpha'_P(0)\simeq 0.15 - 0.25$
(see, for example, \cite{kaid,land,dakhno}). These Pomeron parameters
give for the tensor glueball $M\simeq 1.7 - 2.5$ GeV. \\
 (ii) In the lattice calculations,  a close value was obtained, namely,
$M\simeq 2.2-2.4$ GeV  \cite{lattice}. \\
 (iii) The large width of $f_2(2000)$  can be considered as a
 signature of the glueball origin of this state. Exotic state appearing
 in a set of $q\bar q$ resonances accumulates their widths, thus
 transforming into broad resonance \cite{PR-exotic}.
The phenomenon of width accumulation has been
studied in \cite{APS-PL,AAS-PL} for scalar glueball
$f_0(1200-1600)$, and much earlier this phenomenon  was observed in
nuclear physics \cite{shapiro,okun,stodolsky}.

Direct arguments for the glueball nature of $f_2(2000)$ can be provided
by the relations between decay coupling constants, and for  tensor
glueball such relations were presented  in \cite{glue-2}. In
\cite{Ani,PNPI}, the  extraction of the decay couplings $f_J\to \pi\pi,
\eta\eta, \eta\eta'$ was not performed ---  in the
present paper we fill in
this gap. The $\bar p p \to \pi^0\pi^0, \eta\eta, \eta\eta'$ amplitudes
provide us the following ratios for the $f_2$ resonance couplings,
 $g_{\pi^0\pi^0}:g_{\eta\eta}:g_{\eta\eta'}$ :
\begin{eqnarray}
f_2(1920)\hspace*{0.5cm} &&\ 1:0.56\pm0.08:0.41\pm0.07
\nonumber\\
f_2(2000)\hspace*{0.5cm} &&\
1:0.82\pm0.09:0.37\pm0.22
\nonumber\\
f_2(2020)\hspace*{0.5cm} &&\ 1:0.70\pm0.08:0.54\pm0.18
\nonumber\\
f_2(2240)\hspace*{0.5cm} &&\ 1:0.66\pm0.09:0.40\pm0.14
\nonumber\\
f_2(2300)\hspace*{0.5cm} &&\ 1:0.59\pm0.09:0.56\pm0.17.
\label{2}
\end{eqnarray}
These ratios are to be compared with those given in \cite{glue-2}.

In the leading terms of $1/N_c$-expansion \cite{t'hooft},  there exist
definite ratios for the glueball
decay couplings.
The next-to-leading terms in the decay
couplings give the corrections of the order of $1/N_c$ (see, for
example, \cite{ufn04}); numerical calculations of diagrams tell us that
$1/N_c$ factor leads to a smallness of the order of $1/10$, and we
neglect them.
For the transitions
$ tensor \, glueball\to\pi^0\pi^0,
\eta\eta,\eta\eta'$ the relations in the leading terms of
$1/N_c$-expansion
read (see Table in \cite{glue-2}):
\begin{equation}
 g^{(glueball)}_{\pi^0\pi^0}: g^{(glueball)}_{\eta\eta}:
g^{(glueball)}_{\eta\eta'} =1:(\cos^2\Theta+\lambda\sin^2\Theta)
: (1-\lambda)\sin\Theta\cos\Theta\ .
\label{3}
\end{equation}
Here
$\Theta$ is the mixing angle for  $\eta -\eta'$ mesons:
$\eta=n\bar n \cos\Theta-s\bar s \sin\Theta$ and
$\eta'=n\bar n \sin\Theta+s\bar s \cos\Theta$, where $n\bar n=(u\bar
u+d\bar d)/\sqrt 2$.
We neglect a possible admixture of the gluonium component
in $\eta$ and $\eta'$ (according to \cite{eta-glue}, the gluonium
admixture in $\eta$ is less than 5\%, and in $\eta'$ it is less than
20\%). For the mixing angle $\Theta$ we use $\Theta=37^\circ$.

Suppression parameter $\lambda$ determines
relative production probability of strange quarks by gluon field
$u\bar u:d\bar d:s\bar s=1:1:\lambda$ with $0\leq \lambda \leq 1$.
The data provide us with the following values
of this parameter: $\lambda\simeq 0.5$
\cite{lambda} for  central hadron production in hadron--hadron
high energy collisions,
 $\lambda=0.5-0.8$ \cite{klempt} for the decay of tensor mesons and
 $\lambda=0.5-0.9$ \cite{kmat} for  the decays of $0^{++}$ mesons.

For $(\lambda=0.5$, $\Theta=37^\circ)$ eq.
(\ref{3}) gives us $1:0.82:0.24$, and for
$(\lambda=0.85$, $\Theta=37^\circ)$, correspondingly, $1:0.95:0.07$.
Consequently, the relations between the coupling constants
$g_{\pi^0\pi^0}:g_{\eta\eta}:g_{\eta\eta'}$ for the glueball are to be
as follows:
\begin{equation}
\hspace*{-3cm} 2^{++}glueball \hspace{1cm}
g_{\pi^0\pi^0}:g_{\eta\eta}:g_{\eta\eta'}\ =\
1:(0.82-0.95):(0.24-0.07).
\label{4}
\end{equation}
We see from (\ref{2}) that precisely the coupling
constants of the broad $f_2(2000)$ resonance are inside the
intervals: $0.82\le
g_{\eta\eta}/ g_{\pi^0\pi^0}\le0.95$ and $0.24\ge
g_{\eta\eta'}/g_{\pi^0\pi^0}\ge0.07$ . Hence, it is just this
resonance which can be considered as  tensor glueball,
with $\lambda$ being fixed in the interval $0.5\le\lambda\le0.7$.

Taking
into account that there is no room for $f_2(2000)$ on the
$(n,M^2)$-trajectories \cite{glue-2}, it becomes clear that  this
resonance is indeed the lowest tensor glueball.

The authors are grateful to A.V. Anisovich, D.V. Bugg, L.G. Dakhno,
M.A. Matveev and V.A. Nikonov for useful discussions. The paper was
supported by the grant No. 04-02-17091 of the RFFI.

\begin{figure}
\psfig{file=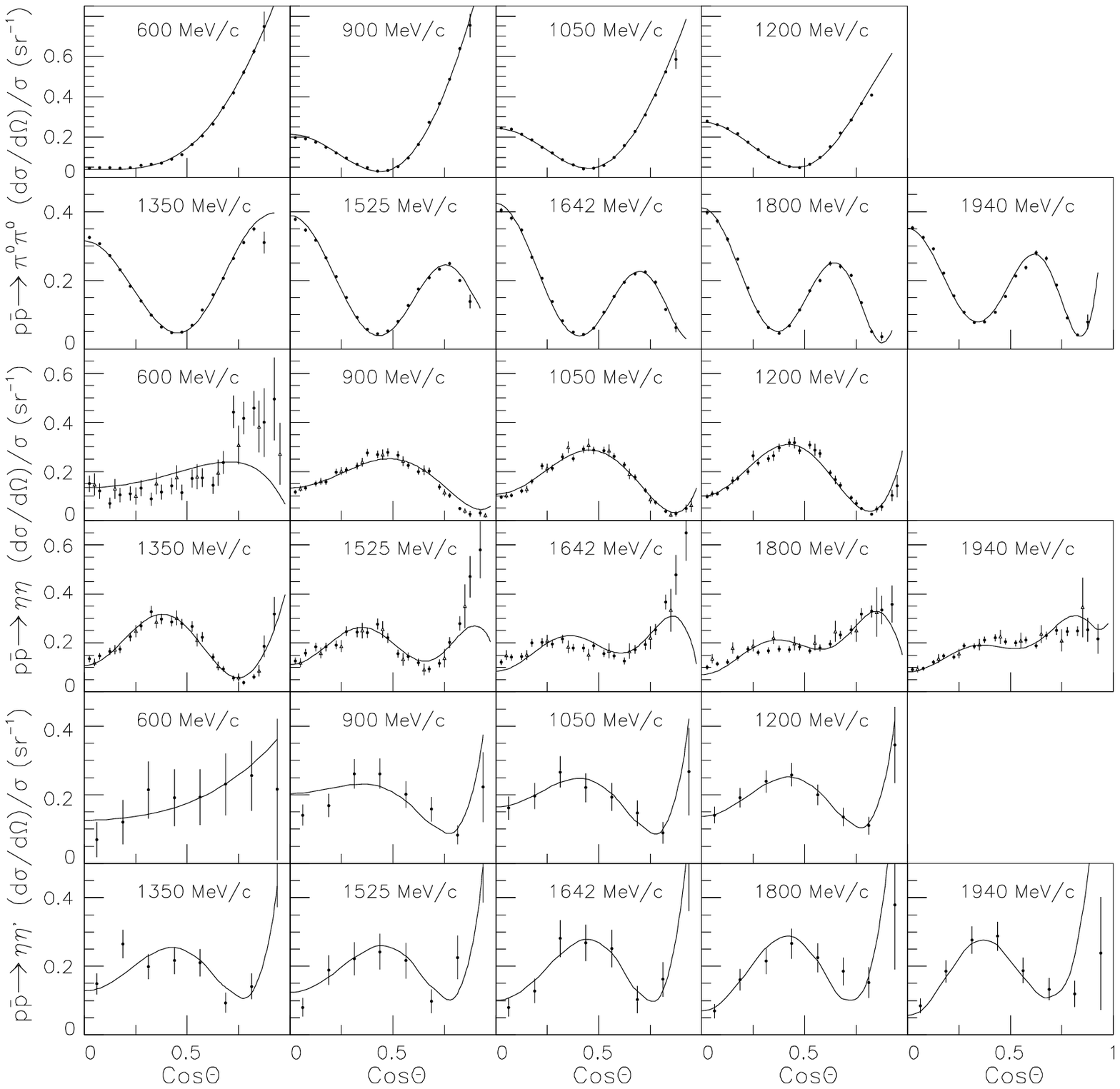,width=15cm}
\caption{ Angle distributions in the reactions
$p\bar p\to \pi\pi,\eta\eta,\eta\eta'$ and their fit to resonances
of eq. (1).}
\end{figure}

\newpage
\begin{figure}
\psfig{file=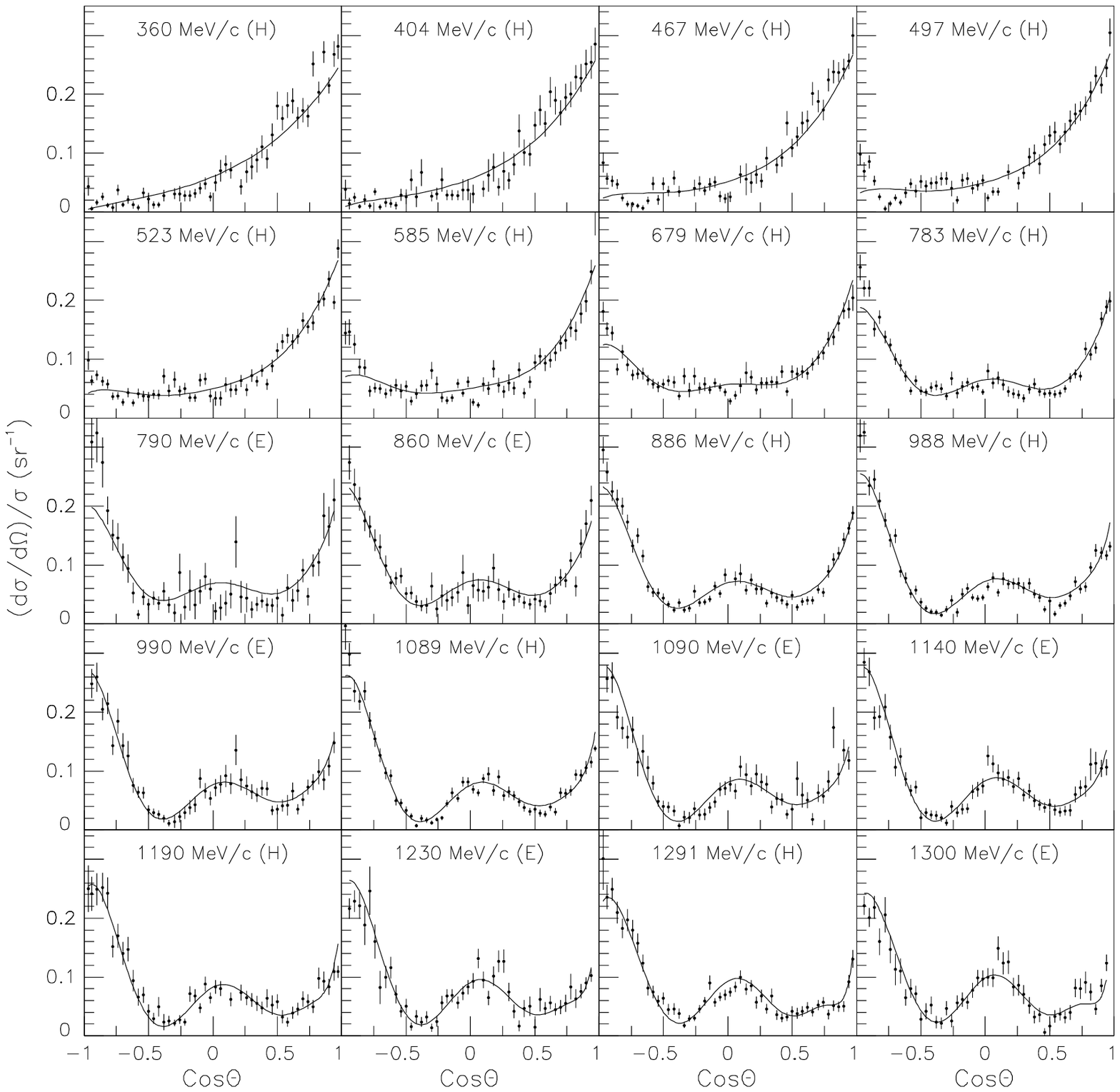,width=15cm}
\caption{
Differential cross sections in the reaction
$p\bar p\to\pi^+\pi^-$  at proton momenta 360-1300 MeV
and their fit to resonances
of eq. (1).}
\end{figure}

\newpage
\begin{figure}
\psfig{file=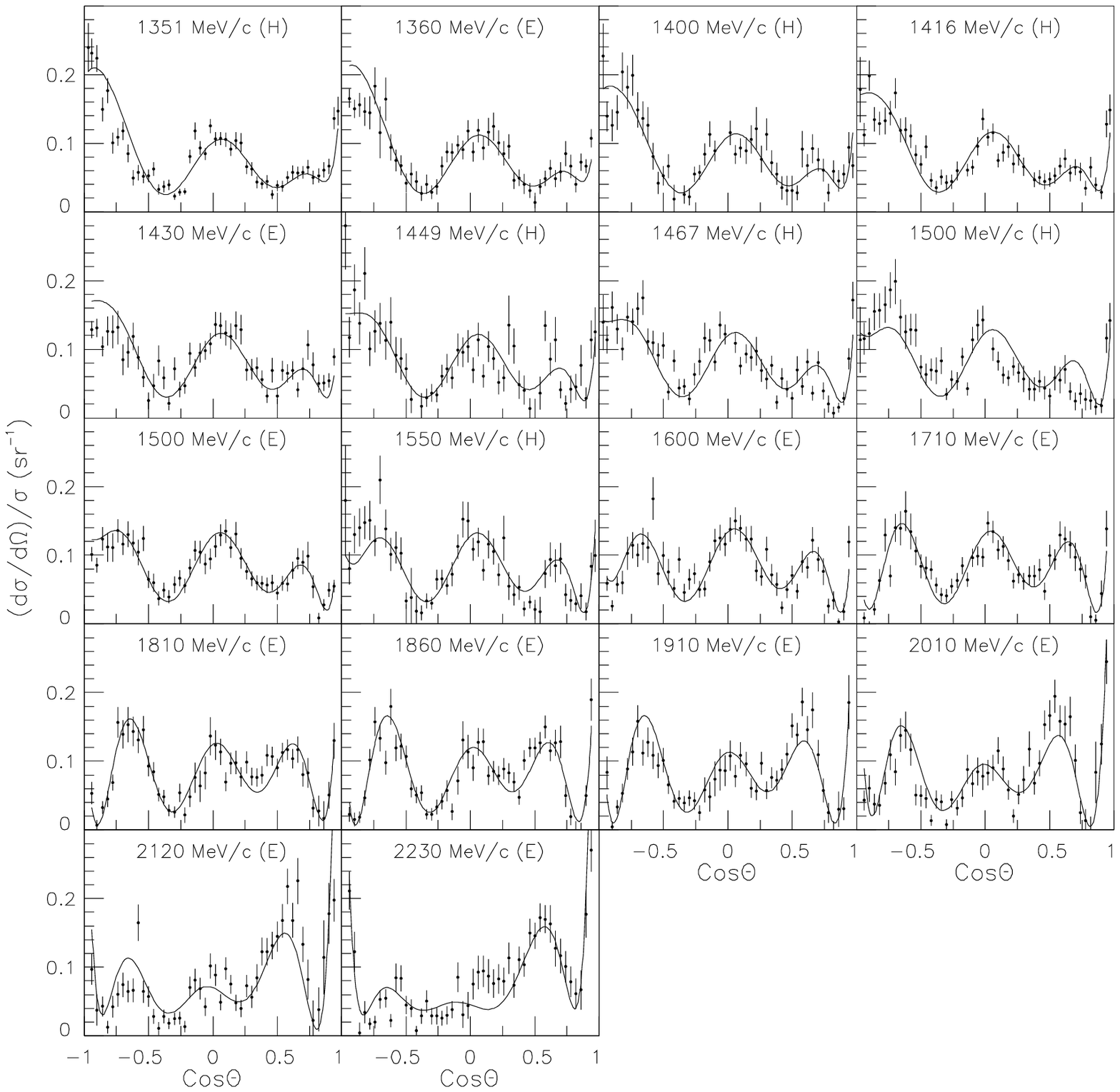,width=15cm}
\caption{
Differential cross sections in the reaction
$p\bar p\to\pi^+\pi^-$  at proton momenta 1350-2230 MeV
and their fit to resonances
of eq. (1).  }
\end{figure}

\newpage
\begin{figure}
\psfig{file=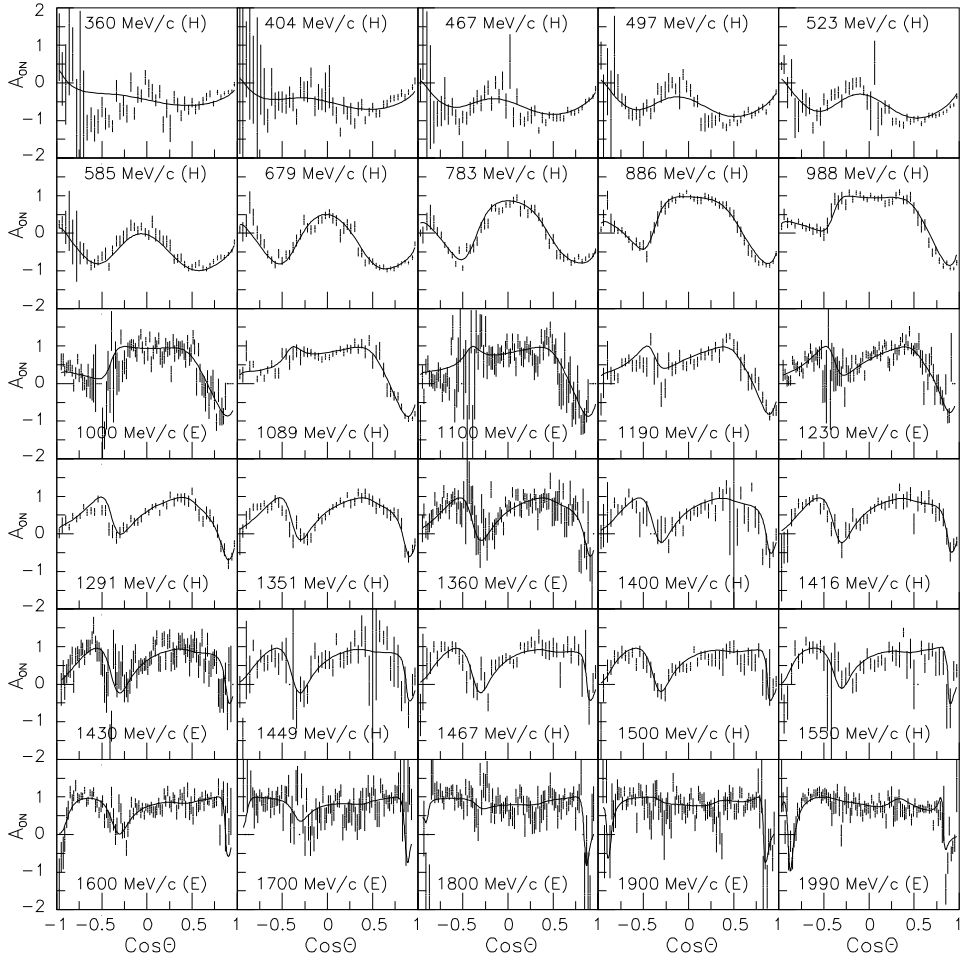,width=15cm}
\caption{
Polarisation in
$p\bar p\to\pi^+\pi^-$
and its fit to resonances
of eq. (1).
}
\end{figure}
\newpage
\begin{figure}
\centerline{\epsfig{file=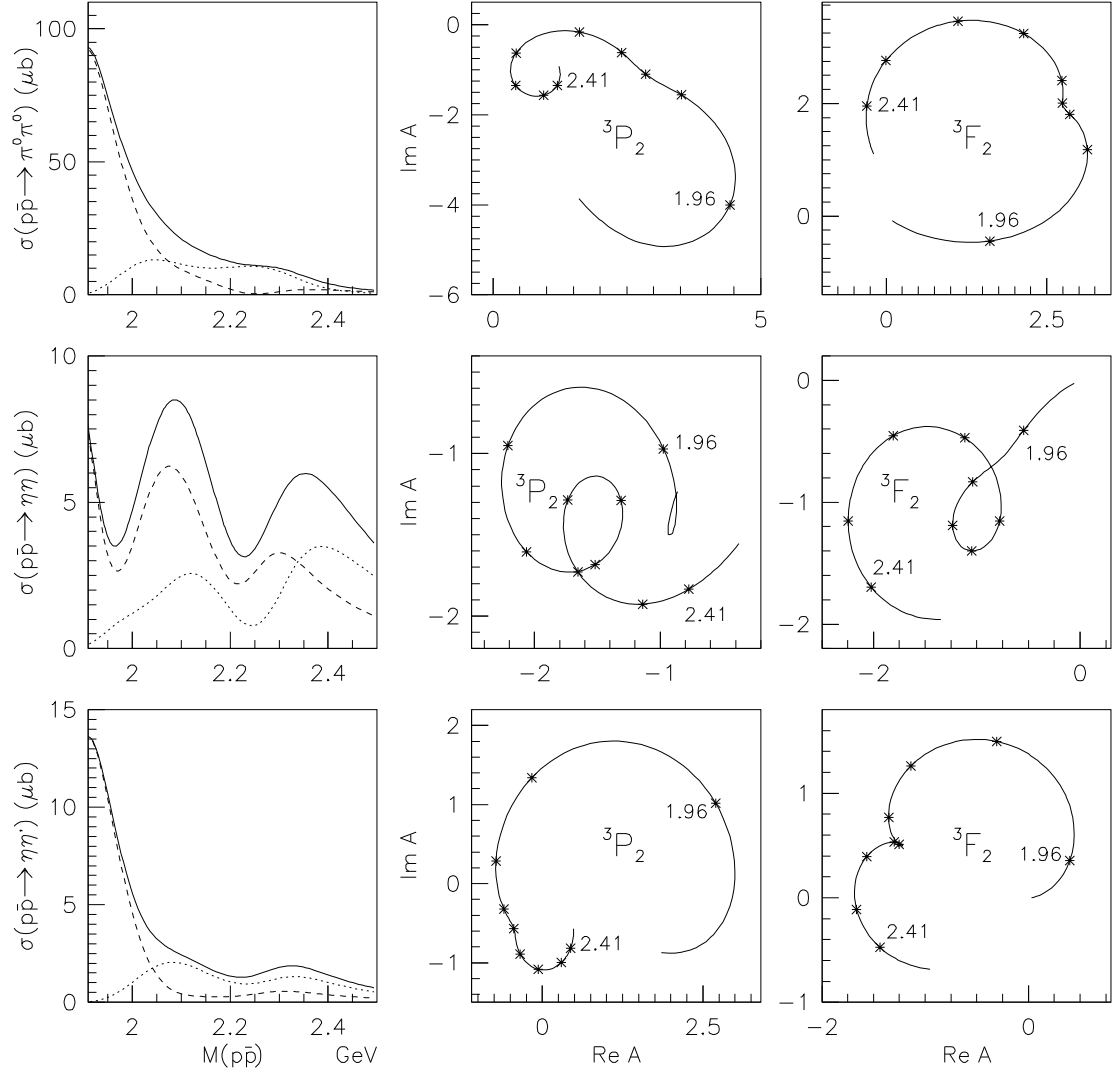,width=17cm}}
\caption{Cross sections and Argand-plots for $^3P_2$ and $^3F_2$ waves
in the reaction $p\bar p\to\pi^0\pi^0,\eta\eta,\eta\eta'$. The upper
row refers to $p\bar p\to\pi^0\pi^0$:
 we demonstrate the cross sections
 for $^3P_2$ and $^3F_2$ waves
(dashed and dotted lines, correspondingly) and  total $(J=2)$ cross
section (solid line) as well as Argand-plots for the $^3P_2$ and
$^3F_2$ wave amplitudes at invariant masses $M=1.962$, $2.050$,
$2.100$, $2.150$, $2.200$, $2.260$, $2.304$, $2.360$, $2.410$ GeV. The
figures on the second and third rows refer to the reactions $p\bar
p\to\eta\eta$  and $p\bar p\to\eta\eta'$.}
\end{figure}

\end{document}